\documentclass[aps,pra,reprint,a4paper]{revtex4-1}
\usepackage[utf8]{inputenc}
\usepackage[english]{babel}
\usepackage{graphicx}
\usepackage{vmargin}
\usepackage{amsmath}
\usepackage{amssymb}
\usepackage{textcomp}
\usepackage{dcolumn}
\newcommand{\BiO}{$\mathrm{BiO}$}
\newcommand{\SiOiv}{\mbox{$\mathrm{SiO}_4$}}
\newcommand{\SiOii}{\mbox{$\mathrm{SiO}_2$}}
\newcommand{\BiiiOiii}{\mbox{$\mathrm{Bi}_2\mathrm{O}_3$}}
\newcommand{\Term}[4]{\mbox{${}^{#1}{#2}_{#3}^{#4}$}}
\newcommand{\gamess}{\mbox{GAMESS~(US)}}
\newcommand{\cminv}{$\textrm{cm}^{-1}$}
\newcommand{\mks}{\textmu{}s}
\begin{document}
\title{%
Interstitial \mbox{BiO} molecule as a center of broadband IR luminescence \\
in bismuth-doped silica glass
}
\author{%
V.O.Sokolov
}
\author{%
V.G.Plotnichenko
}
\author{%
E.M.Dianov
}
\affiliation{%
Fiber Optics Research Center, Russian Academy of Sciences \\
38~Vavilov Street, Moscow, Russia 119333
}

\begin{abstract}
\end{abstract}
\maketitle

\begin{figure}
\includegraphics[scale=0.60, bb=5 0 350 267]{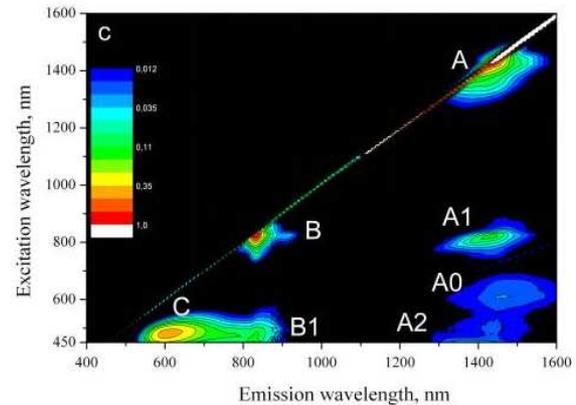}
\caption{%
Luminescence intensity vs. luminescence wavelength and absorption wavelength
at 77~K (figure from Ref.~\cite{Bufetov11b})
}
\label{fig:1}
\end{figure}

In Refs.~\cite{Bufetov11a, Bufetov11b} luminescence spectra have been studied in
optical fibers with silica (\SiOii) glass core doped with bismuth oxide
(\BiiiOiii) only, without any other doping component. Principal results of this
study are shown in Fig.~\ref{fig:1}.

The most characteristic spectral properties found in this study are as follows:
\begin {itemize}
\item[---] the absorption at wavelengths near 1425, 820, 620, and $\lesssim
450$~nm (energy about 7020, 12150, 16130 and $\gtrsim 22200$~\cminv,
respectively) result in IR luminescence with wavelength near 1430~nm (energy
about 6990~\cminv);
\item[---] the absorption at wavelengths of about 820 and $\lesssim 450$~nm
(energy about 12150 and $\gtrsim 22200$~\cminv, respectively) leads to
luminescence at wavelength of about 830~nm (energy about 12050~\cminv);
\item[---] at liquid nitrogen temperatures, weak luminescence at wavelengths of
about 910 and 830~nm (energy about 10990 and 12050~\cminv, respectively), is
excited by absorption at wavelengths of about 820 and 760~nm (energy about 12195
and 13155~\cminv), respectively;
\item[---] lifetimes of the excited states responsible for the luminescence near
1430
and 830~nm are found to be 640 and 40~\mks, respectively.
\end{itemize}

\begin{table*}
\caption{%
Spectroscopic data from Refs.~\cite{Bufetov11a, Bufetov11b} and proposed
interpretation based on transitions in \BiO{} molecule
}
\begin{tabular}{D{.}{}{2}D{.}{}{2}|c|D{.}{}{2}D{.}{}{2}|c}
\hline\hline
\multicolumn{3}{c|}{absorption} & \multicolumn{3}{c}{luminescence} \\[0.4ex]
\hline
&&&&& \\[-2.0ex]
\multicolumn{1}{c}{nm} & \multicolumn{1}{c|}{\cminv} & transition &
\multicolumn{1}{c}{nm} & \multicolumn{1}{c|}{\cminv} & transition \\[0.5ex]
\hline\hline
&&&&& \\[-1.5ex]
  1425. &    7018. &
X$_1$ \Term{2}{\Pi}{1/2}{} $\rightarrow$ X$_2$ \Term{2}{\Pi}{3/2}{} &
  1435. &    6969. &
X$_2$ \Term{2}{\Pi}{3/2}{} $\rightarrow$ X$_1$ \Term{2}{\Pi}{1/2}{} \\
   821. &   12180. &
X$_1^\ast$ \Term{2}{\Pi}{1/2}{} $\rightarrow$ A$_2$ \Term{4}{\Pi}{1/2}{} &
  1430. &    6993. &
X$_2$ \Term{2}{\Pi}{3/2}{} $\rightarrow$ X$_1$ \Term{2}{\Pi}{1/2}{} \\
 < 450. & > 22200. &
X$_1$ \Term{2}{\Pi}{1/2}{} $\rightarrow$ H \Term{2}{\Pi}{1/2}{},
I \Term{4}{\Sigma}{1/2}{-}, etc. &
  1430. &    6993. &
X$_2$ \Term{2}{\Pi}{3/2}{} $\rightarrow$ X$_1$ \Term{2}{\Pi}{1/2}{} \\
   620. &   16129. &
X$_1$ \Term{2}{\Pi}{1/2}{} $\rightarrow$ A$_4$ \Term{4}{\Pi}{1/2}{} (?) &
  1480. &    6757. &
X$_2$ \Term{2}{\Pi}{3/2}{} $\rightarrow$ X$_1^\ast$ \Term{2}{\Pi}{1/2}{} \\
   823. &   12150. &
X$_1$ \Term{2}{\Pi}{1/2}{} $\rightarrow$ A$_2$ \Term{4}{\Pi}{1/2}{} &
   833. &   12005. &
A$_2$ \Term{4}{\Pi}{1/2}{} $\rightarrow$ X$_1$ \Term{2}{\Pi}{1/2}{} \\
 < 450. & > 22200. &
X$_1$ \Term{2}{\Pi}{1/2}{} $\rightarrow$ H \Term{2}{\Pi}{1/2}{},
I \Term{4}{\Sigma}{1/2}{-}, etc. &
   830. &   12048. &
A$_2$ \Term{4}{\Pi}{1/2}{} $\rightarrow$ X$_1$ \Term{2}{\Pi}{1/2}{} \\
   820. &   12195. &
X$_1$ \Term{2}{\Pi}{1/2}{} $\rightarrow$ A$_2$ \Term{4}{\Pi}{1/2}{} &
   910. &   10990. &
A$_2$ \Term{4}{\Pi}{1/2}{} $\rightarrow$ X$_1^\ast$ \Term{2}{\Pi}{1/2}{} \\
   760. &   13158. &
X$_1$ \Term{2}{\Pi}{1/2}{} $\rightarrow$ A$_2^\ast$ \Term{4}{\Pi}{1/2}{} &
   830. &   12048. &
A$_2$ \Term{4}{\Pi}{1/2}{} $\rightarrow$ X$_1$ \Term{2}{\Pi}{1/2}{} \\[0.5ex]
\hline\hline
\end{tabular}
\label{tab:spectra}
\par
\begin{flushleft}
\begin{footnotesize}
${}^\ast)$~absorption or luminescence bands wavelength corresponds to
transitions to or from both the ground and excited vibrational states of the
the ground electronic term (X$_1$ \Term{2}{\Pi}{1/2}{}) of \BiO{} molecule
\end{footnotesize}
\end{flushleft}
\end{table*}

\begin{figure}
\includegraphics[scale=0.44, bb=75 295 545 770]{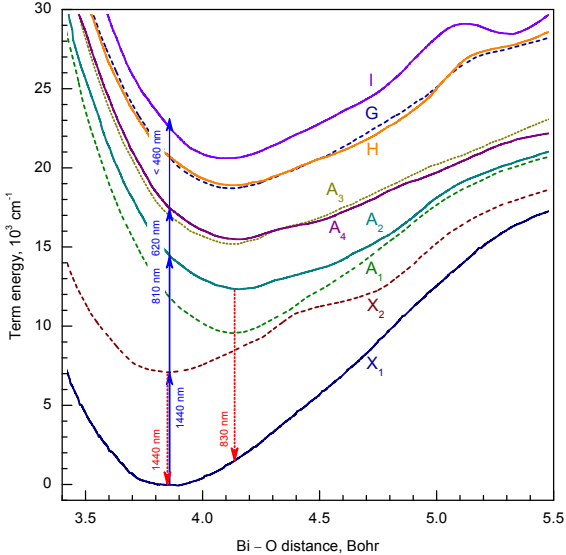}
\caption{%
Total energy curves of the lower electronic states of \BiO{} molecule according
to calculation \cite{Alekseyev94} and experiment \cite{Shestakov98} (solid
lines: $\Omega = 1/2$ states; dashed lines: $\Omega = 3/2$ states; dotted
lines: $\Omega = 5/2$ states) and the transitions corresponding to the
absorption (solid arrows) and luminescence (dashed arrows) bands measured in
Refs.~\cite{Bufetov11a, Bufetov11b}
}
\label{fig:2}
\end{figure}
The described absorption and luminescence features of the bismuth-doped silica
glass have engaged our attention because of close agreement with well-known
spectral properties of bismuth monoxide molecule, \BiO{} \cite{Shestakov98,
DiatomicMolecules, Alekseyev94}.

The Table~\ref{tab:spectra} gives the experimental data of
Refs.~\cite{Bufetov11a, Bufetov11b} and their interpretation based on
spectroscopic data available for \BiO{} molecule \cite{Shestakov98, Alekseyev94,
DiatomicMolecules}. Fig.~\ref{fig:2} shows the total energy curves of the \BiO{}
molecule constructed using the configuration interaction calculations with
spin-orbit interaction taken into account \cite{Alekseyev94} corrected to
achieve more close agreement with the experimental data \cite{Shestakov98,
DiatomicMolecules}.

\BiO{} molecule is known to have low-lying long-living excited electronic state,
X$_2$~\Term{2}{\Pi}{3/2}{}, with energy of about 7090~\cminv{} and lifetime
corresponding to transition to the ground state of $480 \pm 100$~\mks{}
\cite{Shestakov98} (calculated values are 6810~\cminv{} and 2700~\mks,
respectively \cite{Alekseyev94}). It should be emphasized that the minima of the
excited-state term, X$_2$, and the ground-state term,
X$_1$~\Term{2}{\Pi}{3/2}{}, practically coincide (Fig.~\ref{fig:2}) and hence
the $\textrm{X}_2 \rightarrow \textrm{X}_1$ luminescence excited in
$\textrm{X}_1 \rightarrow \textrm{X}_2$ absorption should be very small.

The X$_2$ excited state can be populated not only in this absorption, but in
non-radiative transitions from higher-lying electronic states as well. The most
intense characteristic absorption bands of \BiO{} molecule correspond to
transitions from the ground electronic state, X$_1$, to several excited
electronic states, namely, to A$_2$~\Term{4}{\Pi}{1/2}{}{} state (transition
energy about 12200~\cminv{} or wavelength about 820~nm corresponds to averaged
energy of 0--0, 1--0, 2--0, and 3--0 transitions from vibrational states of the
X$_1$ term to the ground vibrational state of the A$_2$ with Bi$\relbar$O
stretching vibration frequency being approximately 690~\cminv{} in the X$_1$
state \cite{Shestakov98, DiatomicMolecules}), to H~\Term{2}{\Pi}{1/2}{}{} and
I~\Term{4}{\Sigma}{1/2}{-}{} terms (transition energy 20500--22000~\cminv{} or
wavelength 450--490~nm), and to a number of higher-lying electronic states
(transition energy $\gtrsim 25000$~\cminv{} or wavelength $\lesssim 400$~nm).
The calculations described in Ref.~\cite{Alekseyev94} also predict an absorption
band with energy about 16130~\cminv{} (wavelength about 630~nm) caused by
transition from the X$_1$ ground state to the A$_4$~\Term{4}{\Pi}{1/2}{}{}
state. Lifetime of the A$_4$~\Term{4}{\Pi}{1/2}{}{} excited state corresponding
to transition to the ground state is found to be about 165~\mks{}
\cite{Alekseyev94}. When \BiO{} molecule is excited to one of these states,
non-radiative transitions to the first excited state, X$_2$ with subsequent
$\textrm{X}_2 \rightarrow \textrm{X}_1$ luminescence would be expected. On the
other hand, radiative transitions accompanied by luminescence are possible as
well. The $\textrm{A}_2 \rightarrow \textrm{X}_1$ transition is most likely to
occur (energy about 12005~\cminv, wavelength about 835~nm, the A$_2$ state
lifetime $9 \pm 2$~\mks{} \cite{Shestakov98}). In this case the Stokes shift is
also small owing to significant contribution of vibrational excitations of the
ground electronic state.

Mention should be made of noticeably longer lifetimes of the states responsible
for the luminescence near 1430 and 830~nm obtained in experiments with fibers
\cite{Bufetov11a} (see above) in comparison with both other experimental data
\cite{Shestakov98, DiatomicMolecules} and results of calculations
\cite{Alekseyev94}. In general, this could be explained by significant influence
of re-absorption under strong overlap of absorption and luminescence bands
enhanced due to guiding effect of the fiber. Indeed, in Ref.~\cite{Bufetov11a}
the luminescence was excited in the absorption band near 820~nm when measuring
lifetimes.

The weak luminescence near 910~nm (10990~\cminv) excited by absorption at about
820~nm (12195~\cminv) may be attributed to transitions between the X$_1$ and
A$_2$ states. The absorption corresponds to the above-described case and the
luminescence corresponds to transitions from the ground vibrational state of
the A$_2$ electronic term to several excited vibrational states of the X$_1$
term, mainly 0--1, 0--2, and 0--3 transitions. Similarly, the weak luminescence
near 830~nm (12050~\cminv) excited by absorption at about 760~nm (13155~\cminv)
could be explained by transitions between the X$_1$ and A$_2$ states. Now
absorption occurs from the ground vibrational state of the X$_1$ electronic term
to several excited vibrational states of the A$_2$ term (mainly 0--1, 0--2, and
0--3 transitions, with the vibration frequency of the A$_2$ term being about
505~\cminv{} \cite{Shestakov98, Alekseyev94}) and the luminescence corresponds
to the above-described case.

Taken together, the foregoing considerations suggest that the absorption and
IR luminescence in bismuth-doped silica glass observed in
Refs.~\cite{Bufetov11a, Bufetov11b} are not unlikely to be caused by transitions
in interstitial \BiO{} molecules.

To make sure that interstitial \BiO{} molecule can actually occur in the silica
glass network, we performed quantum chemical modeling based on cluster models.
The clusters contained one or two sixfold rings formed by \SiOiv{} tetrahedra.
In the second case two coaxial rings were located at a distance of 3~\AA{} from
each other. Dangling bonds of the outer oxygen atoms of the clusters were
saturated by hydrogen atoms to model the surrounding network. \BiO{} molecule
was initially placed in the central part of each cluster on the axis of the ring
(or rings), and then complete geometry optimization was performed. All
calculations were accomplished with \gamess{} program \cite{gamess} by density
functional method using BLYP functional, which is known to provide good
agreement between calculated and experimental geometric parameters. Bases and
effective core potentials developed in \cite{SBKJC} were used. For oxygen and
bismuth atoms an additional $d$-type polarization function was included in the
basis ($\zeta_{\textrm{O}} = 0.800$~Bohr, $\zeta_{\textrm{Bi}} = 0.185$~Bohr).
Standard \hbox{3-21G} basis was used for hydrogen atoms.

\begin{figure}
\begin{center}
\includegraphics[scale=0.275, bb=0 0 750 770]{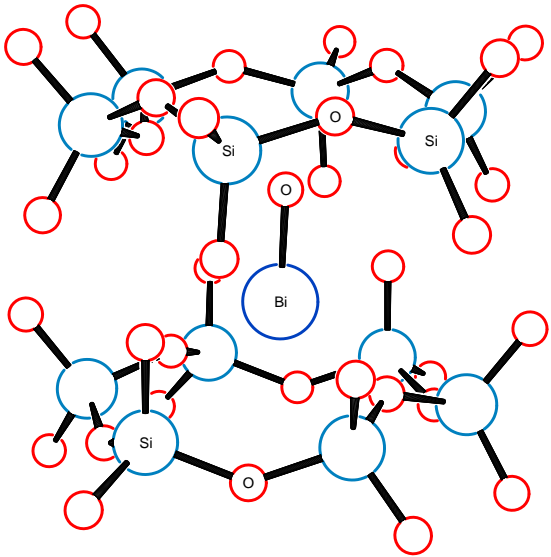}
\end{center}
\vspace{1.0\baselineskip}
\caption{%
\BiO{} molecule in ring interstitial of silica glass network according to our
calculations
}
\label{fig:3}
\end{figure}
Notice that no stable interstitial position has been found for \BiO{} molecule
in alumosilicate glass in similar calculations performed recently
\cite{Sokolov09}: \BiO{} molecule is found to react with the surrounding
oxygen atoms with threefold coordinated bismuth atom formed as a result.

However in the case of pure silica there is an equilibrium position of \BiO{}
molecule in \SiOii{} network interstitial formed by sixfold rings: \BiO{}
molecule takes position between two rings being aligned along the rings axis
(Fig.~\ref{fig:3}). Such a position turns out to be quite stable: if the
molecule is forced either to be deviated from the rings axis or to be shifted
apart from the center of the rings, no reaction of above-mentioned type occurs
with the environment and the molecule returns to its equilibrium position. Thus,
our calculations favor the assumption under consideration.

Calculations of the vibrational properties of \BiO{} molecule in ring
interstitial site of \SiOii{} glass network showed that the frequency of the
Bi$\relbar$O stretching vibration is close to the free molecule ($\sim
700$~\cminv). This vibration could be observable both in IR absorption
(absorption coefficient $\sim 1.6$~Db$^2/\textrm{amu}/\textrm{\AA}^2$), and in
Raman scattering (Raman intensity $\sim 32$~\AA$^4/\textrm{amu}$, the
depolarization coefficient $\approx 0.015$). In addition, there are librational
(395 and 430~\cminv) and translational (60, 80, and 95~\cminv) vibrations of
\BiO{} molecule in the interstitial site, but both IR absorption and Raman
scattering corresponding to those seem to be too low to be observed in
experiment, taking into consideration low concentrations of bismuth in the
doped glasses.

From the above reasoning, several experimental studies of bismuth-doped silica
seem to be of importance, as follows:
\begin{enumerate}
\item measure lifetimes of the states responsible for luminescence near 1430 and
830~nm with the luminescence being excited in the short-wavelength absorption
band, $\lesssim 450$~nm, to eliminate re-absorption;
\item measure Raman spectra in fibers with bismuth-doped silica glass core to
detect the Raman band corresponding to stretching vibrations of \BiO{} molecule;
\item measure absorption and luminescence high-resolution spectra at low
temperatures ($T \lesssim 20$~K is estimated to be enough assuming translational
vibrations of interstitial \BiO{} molecules to contribute predominantly to the
bands broadening) to detect vibrational structure of both absorption and
luminescence bands;
\item repeat both the experiments of Refs.~\cite{Bufetov11a, Bufetov11b} and
the above-enumerated experiments with bismuth-doped silica crystals (quartz,
cristobalite, or tridymite).
\end{enumerate}

%
%

\end{document}